\documentclass[twocolumn,showpacs,preprintnumbers,amsmath,amssymb]{revtex4}

\usepackage{graphicx}
\usepackage{dcolumn}
\usepackage{bm}

\begin{document}

\title{ANTIPROTON ABSORPTION IN NUCLEI}

\author{Yu.T.Kiselev}
\email{yurikis@itep.ru}
  \affiliation{Institute for Theoretical and Experimental Physics,
               Moscow, Russia }

\author{E.Ya.Paryev}
  \affiliation{Institute for Nuclear Research, Russian Academy of
               Sciences, Moscow, Russia}%

\date{\today}

\begin{abstract}

We present the analysis of experimental data on forward antiproton
production on nuclei. The calculations are done in the framework
of a folding model which takes properly into account both
incoherent direct proton-nucleon and cascade pion-nucleon
antiproton production processes as well as internal nucleon
momentum distribution. The effective antiproton-nucleon cross
section in nuclear matter and the imaginary part of the antiproton
nuclear optical potential are estimated to be $25-45$~mb and
$-(38-56)$~MeV at normal nuclear matter density, respectively. The
results of the performed analysis evidence for the decreasing of
antiproton absorption in the nuclear medium.
\end{abstract}

\maketitle

\section{\label{sec:level1}Introduction}

The modification of the hadron properties and interactions in
baryon environment is one of the important topics of contemporary
strong interaction physics. Such a modification has been predicted
within various theoretical approaches such as QCD sum rules \cite{Ref1},
chiral dynamics \cite{Ref2}, relativistic mean-field \cite{Ref3}
and quark-meson coupling models \cite{Ref4}. Embedding a hadron in
nuclear matter can cause a change of its mass and width. An
evidence for a decrease of the ${\rho}$ meson in-medium mass in
heavy-ion collisions was obtained by the CERES collaboration at
CERN \cite{Ref5} and later indicated by the STAR experiment at RHIC
\cite{Ref6}. Very recently, the CBELSA/TAPS collaboration
reported on the first observation of sizable lowering of the ${\omega}$
meson mass at normal nuclear matter density in the ${\gamma}Nb$
reaction \cite{Ref7}.

The interaction of a hadron in nuclear medium is expected to be
also modified leading, in particular, to change of its absorption
in nucleus. The results of recent experiment \cite{Ref8} on the
$\phi$ meson photo-production from different nuclei evidence for
significant enhancement of the $\phi$ $-$ nucleon cross section in the
nuclear environment. The study of the propagation of antiprotons
in nuclear matter can provide the valuable information on the
in-medium modification of the $\bar p$ $-$ nucleon interaction.

Anomalous weak absorption of the antiprotons produced in
proton-nucleus collisions had been first observed in \cite{Ref9}
and then was confirmed in \cite{Ref10,Ref11}. The inclusive and
semi-inclusive $\bar p$ production in $p$Be, $p$Cu and $p$Au
interactions was studied at AGS energies of 12.3 and 17.5 GeV
\cite{Ref12}. The analysis within the Glauber-type multiple
scattering model performed in \cite{Ref12} shown that the $\bar p$
absorption cross section is at least a factor of 5 smaller than
the free $\bar p N$ annihilation cross section. Several mechanisms
for the decreased $\bar p$ absorption have been proposed including
finite antiproton formation time \cite{Ref9,Ref12,Ref13},
''shielding'' of the absorption process \cite{Ref14}, strong
reduction of the phase space available for annihilation due to
drop of in-medium $\bar p$ and nucleon masses \cite{Ref15}, the
formation of $\bar p p$ bound state \cite{Ref16}, which was
apparently observed in \cite{Ref17}. All of such arguments
manifest themselves as a suppression of the antiproton
annihilation with nuclear nucleons. Proton-nucleus collisions
provide a suitable tool for disentangling of the suggested
theoretical schemes and elucidating the dynamics of in-medium
antiproton production and absorption in nuclear environment. Using
the nuclei as an absorbers has an advantage to have a well
defined, nearly static density profile. In this work we present
the analysis of inclusive antiproton production in proton-nucleus
collisions aimed at the evaluation of the effective
antiproton-nucleon cross section in nuclear matter and the
imaginary part of the $\bar p$ - nuclear potential.

\section{\label{sec:level1}Experimental data}
Recently, the differential cross sections for antiproton
production on Be, Al, Cu and Ta targets were measured at 10 GeV
ITEP synchrotron \cite{Ref18}. The data were taken at initial
kinetic proton energies of 5.5, 7.2 and 9.2 GeV. Secondary
antiprotons in the momentum range $0.73-2.47$ GeV/c were detected at $10.5^o$ (lab).

For the evaluation of the antiproton absorption in nuclear matter
we use the subset of the experimental data which includes 20
values of the measured differential cross sections on Cu and Ta
targets at projectile energies of $7.2$ and $9.2$ GeV. The reasons
for that are the following. First, evidently that the absorption
effect is most important for middle and heavy nuclei. Second, it
is known that the hadron production in proton-induced reactions at
near threshold energies can be described as a superposition of
individual collisions of the projectile protons and secondary
pions with target nucleons, participating in the Fermi motion. In
such a case the elementary hadron production can be described by the phase
space calculations normalized by the corresponding total cross
sections. The experimental information about the behavior of the
total cross sections for antiproton production in the elementary
$pN$ and $\pi N$ collisions is available at the excess of
collision energy $\Delta=\sqrt{S}-\sqrt{S_0}\ge 1$ ~GeV above the
corresponding thresholds. The main contribution to the cross
section for antiproton production on nuclei comes from the range
of $\Delta >0.5$ GeV at initial proton energy of $7.2$ GeV and
from $\Delta>0.8$ GeV at $9.2$ GeV which are close to the
experimentally studied range of $\Delta$. That ensures the rather
reliable using of the elementary $\bar p$ production cross
sections from \cite{Ref19,Ref20} in our calculations of antiproton
creation on nuclei at beam energies of $7.2$ and $9.2$ GeV. Third,
according to the study \cite{Ref20}, the influence of the real
part of the $\bar p$-nuclear optical potential on the subthreshold
antiproton production in $pA$ collisions diminishes gradually with
increasing bombarding energy and becomes insignificant already at
threshold energy of 5.6 GeV. Finally, at above threshold energies
the internal nucleon momentum distribution can be safely used
instead of the nucleon spectral function which simplifies the
calculations of $\bar p$ production in pA-reactions.

\section{\label{sec:level1}Folding model}

The obtained data were analyzed in the framework of the folding
model [see ~\cite{Ref21} for details] accounting for the
elementary processes which have the lowest free production
thresholds. In addition to the direct antiproton creation via the
first chance inelastic proton-nucleon collision, $p+N \rightarrow
\bar p + X$, it is also possible to produce $\bar p$ in two-step
cascade process $p+N \rightarrow N+N+ \pi$ followed by
$\pi+N\rightarrow \bar p+X$. Although the threshold for the latter
is lower, only high momentum pions contribute to the cross section.

The invariant inclusive cross section for the production on
nucleus with atomic mass A at small laboratory angles an antiproton
with the total energy $E_{\bar p}$ and
momentum $\vec P_{\bar p}$ via the direct reaction channel can be
represented as follows:
\begin{widetext}
 \begin{eqnarray}
    E_{\bar p}\frac{d\sigma_{pA\to{\bar pX}}^{dir}(\vec {P_0})}{d\vec P_{\bar p}}=
    \left\{
     2\pi A \int \limits_0^{\infty} bdb
            \int \limits_{-\infty}^{\infty} dz
     \rho(\sqrt{z^2+b^2})
     \exp
     \left(\
     -\sigma_{pN}^{in}A
            \int \limits_{-\infty}^{z} \rho(\sqrt{b^2+x^2})dx
     -\sigma_{\bar pN}^{eff} A
            \int \limits_{z}^{\infty}\rho(\sqrt{b^2+x^2})dx
      \right)\
    \right\}
\nonumber \\
      \times \left\{
            \int d\vec q n(\vec q)E_{\bar p}
            \frac{d \sigma_{pN\to{\bar pX}}(\sqrt S, \vec P_{\bar p})}{d \vec P_{\bar p}}
    \right\}.
    \label{eq:a}
 \end{eqnarray}
\end{widetext}

The expression in the first braces describes the distortion of the
incident proton and the  absorption of the outgoing antiproton in
its way out of nucleus. Here $b$ and $z$ stand
for the impact parameter and the $z-$component of the coordinate
along the beam axis, respectively. We use $\sigma_{pN}^{in}$ =30
mb for both considered projectile proton energies which is assumed to be the
same for the pp- and pn-collisions. In Eq.(~\ref{eq:a})
$\sigma_{\bar p N}^{eff}$ stands for the effective in-medium
antiproton-nucleon cross section. Any difference between the
antiproton absorption on nuclear protons and neutrons is
disregarded.

In the calculation of $\bar p$ production on Cu and Ta targets we
use for the density $\rho(\vec r)$ the two-parameter Fermi distribution 
normalized to unity
with parameters R=4.20 fm (Cu), R=6.16 fm (Ta), a=0.55 fm for
both nuclei and a nuclear saturation density $\rho_0$=$0.17$ ~$fm^{-3}$.

The expression in the second braces connects the cross section for
$\bar p$ production in the collision of incident proton with
off-shell struck target nucleon to the elementary on-shell cross section.

The total energy $\omega$ of the struck nucleon is related
to its momentum $\vec q$ as follows:
\begin{equation}
    \omega=m-\frac{{\vec q}^2}{2M_{A-1}} - \epsilon,
    \label{eq:2}
\end{equation}
where $m$ and $M_{A-1}$ stand for free nucleon mass and the mass
of the recoiling target nucleus in its ground state, respectively.
The nucleon binding energy $\epsilon$ is taken equal to 6 MeV.

 The invariant collision energy squared is read:
\begin{equation}
    S = (E_0+\omega)^2 - (\vec P
    _0+\vec q)^2,
    \label{eq:3}
\end{equation}
where $E_0$ and $\vec P_0$ are total energy and momentum of the
projectile proton, respectively.

The internal nucleon momentum distributions $n(\vec q)$ in the
target nuclei are assumed to be in the form:
\begin{eqnarray}
    n(\vec q)= \frac{1}{(2\pi)^{3/2}(1+h)} \nonumber\\
 \times
 \left[\
   \frac{1}{{\sigma_1}^3}\exp(\frac{{-\vec q^2}}{2{\sigma_1}^2})+
   \frac{h}{{\sigma_2}^3}\exp(\frac{{-\vec q^2}}{2{\sigma_2}^2})
 \right]\,
 \label{eq:4}
\end{eqnarray}
with parameters $h$, $\sigma_1$ and $\sigma_2$ derived from the
data on proton-induced subthreshold and near threshold $K^+$
production on the same set of nuclear targets \cite{Ref22}. The
second term in the square brackets represents the high momentum
component of the internal nucleon momentum distributions $n(\vec
q)$ which plays a minor role at considered incident proton energies
of 7.2 and 9.2 GeV.

As was above mentioned, in our approach the invariant inclusive
cross section for antiproton production in the elementary $p+N
\rightarrow \bar p + X$ reaction has been described by the
four-body phase space calculations normalized to the corresponding
total cross section:
 \begin{eqnarray}
    E_{\bar p}
       \frac{d\sigma_{pN \to\bar pX}(\sqrt S,\vec P_{\bar p})}{d\vec P_{\bar
       p}}=  \nonumber\\
    \sigma_{pN\to\bar pX}^{tot}(\sqrt S)
       \frac{R_3(S_{NNN},m,m,m)}{2R_4(S,m_{\bar p},m,m,m)}.
       \label{eq:5}
 \end{eqnarray}
Here $R_3$ and $R_4$ are three-body and four-body phase spaces of
the reaction $p+N\rightarrow \bar p+N+N+N$, while $S_{NNN}$
denotes the invariant energy squared of the not detectable
three-nucleon system.

The invariant inclusive cross section for the antiproton
production on nucleus A at small laboratory angles via the pion induced
reaction channel can be represented as follows~\cite{Ref21}:
\begin{widetext}
  \begin{eqnarray}
   E_{\bar p}
     \frac{d{\sigma_{pA \to \bar pX}}^{casc}(\vec P_0)}{d\vec P_{\bar p}}=
     \frac{I_V[A,\sigma_{pN}^{in},\sigma_{\pi N}^{tot},\sigma_{\bar pN}^{eff},0^{\circ}]}
          {I'_V[A,\sigma_{pN}^{in},\sigma_{\pi N}^{tot},0^{\circ}]}
   \times \nonumber \\
     \sum_{\pi^+,\pi^-,\pi^0}\int d
     \bigl
       (\frac{\vec P_{\pi}}{P_{\pi}}
     \bigr)
     \int {P_{\pi}}^2 dP_{\pi}
       \frac{d\sigma_{pA\to\pi X}(P_{\pi})}{d\vec P_{\pi}}
     \int d\vec q n(\vec q)
     \langle
          E_{\bar p}
       \frac{d \sigma_{\pi N\to\bar pX}(\sqrt S_{\pi},\vec P_{\bar p})}{d\vec P_{\bar
       p}}
    \rangle.
       \label{eq:6}
  \end{eqnarray}
\end{widetext}

The first factor in Eq.~\ref{eq:6} represents the ratio of the
survival probability of an antiproton ($I_V$) to that of a pion
($I'_V$):
\begin{widetext}
  \begin{eqnarray}
     I_V=
     2\pi A^2\int \limits_{0}^{\infty} bdb
               \int \limits_{-\infty}^{\infty} dz \rho(\sqrt{b^2+z^2})
               \int \limits_{0}^{\infty} dl \rho (\sqrt{b^2+(z+l)^2})
      \exp[(
                -\sigma_{pN}^{in} A
                   \int\limits_{-\infty}^z \rho(\sqrt{b^2+x^2})dx -
                   \nonumber\\
                \sigma_{\pi N}^{tot} A
                   \int\limits_z^{z+l} \rho (\sqrt{b^2+x^2})dx -
                \sigma_{\bar pN}^{eff} A
                   \int\limits_{z+l}^{\infty}\rho(\sqrt{b^2+x^2})dx
             ) ] \nonumber\\
        I'_V  = 2\pi A  \int \limits_0^{\infty} bdb
               \int \limits_{-\infty}^{\infty} dz \rho(\sqrt{b^2+z^2})
             \exp[-\sigma_{pN}^{in} A
                      \int \limits_{-\infty}^z\rho(\sqrt{b^2+x^2})dx-
                   \sigma_{\pi N}^{tot} A
                      \int \limits_z^{\infty} \rho (\sqrt{b^2+x^2})dx].
  \label{eq:7}
  \end{eqnarray}
\end{widetext}
In the following calculations $\sigma_{\pi N}^{tot} = 35$ mb for
all pion momenta.

The expression for the cross section for $\bar p$ production in
the elementary $\pi N$ collisions has the form similar to that for
proton-induced reaction:
\begin{eqnarray}
\langle E_{\bar p}
        \frac {d \sigma_{\pi N \to \bar pX}(\sqrt{S_{\pi}},\vec P_{\bar p})}
              {d\vec P_{\bar p}}
 \rangle =
       \frac{Z}{A}
       E_{\bar p}\frac {d \sigma_{\pi p \to {\bar pX}}(\sqrt{S_{\pi}},\vec P_{\bar p})}
             {d\vec P_{\bar p}} \nonumber \\ +
       \frac {A-Z}{A}
    E_{\bar p} \frac {d\sigma_{\pi n\to \bar pX}(\sqrt{S_{\pi}},\vec P_{\bar p})}
             {d\vec P_{\bar p}},
\end{eqnarray}
where Z stands for the atomic weight of the nucleus and
  \begin{eqnarray}
E_{\bar p}
 \frac{d\sigma_{\pi N\to \bar pX}(\sqrt{S_{\pi}},\vec P_{\bar p})}
      {d\vec P_{\bar p}}=
 \nonumber\\
  \frac{\pi}{4} \sigma_{\pi N\to \bar pX}^{tot}(\sqrt{S_\pi})
  \frac{\sqrt{S_{NN}-4m^2}}
       {R_3(S_{\pi},m_{\bar p},m,m)\sqrt{S_{NN}}}.
  \label{eq:8}
  \end{eqnarray}
Here $R_3$ is the three-body phase space of the reaction
$\pi+N\rightarrow \bar p+N+N$, while $S_{NN}$ denotes the
invariant energy squared of the not detectable two-nucleon system. In
case of pion-induced antiproton production the invariant collision
energy squared is read:
\begin{equation}
 S_{\pi} = (E_{\pi}+\omega)^2 - (\vec P_{\pi}+\vec q)^2,
 \label{eq:9}
\end{equation}
where $E_{\pi}$ and $\vec P_{\pi}$ are the pion total energy and
momentum, respectively.

The elementary $\bar p$ production processes $\pi^+ n \to p\bar p
p$, $\pi^0 n \to p\bar p n$, $\pi^0 p \to p \bar p p$, $\pi^- p
\to p\bar p n$ and $\pi^- n \to n\bar p n$ have been included in
our calculations of the $\bar p$ creation on nuclei of interest.

The differential cross sections for pion production by protons on
nuclear targets were deduced from data \cite{Ref10,Ref23,Ref24}.
High momentum part of the pion spectra is approximated in the
form:
 \begin{eqnarray}
E_{\pi}
  \frac{d\sigma_{pA\to\pi X}(\vec P_{\pi})}{d\vec P_{\pi}} =
\nonumber\\
   A(1-X_F^R)^{B-(X_F^R)^2+CP_{\bot}^2} ~[mb ~GeV^{-2}c^{3}sr^{-1}]
   \label{eq:10}
 \end{eqnarray}
 with constants A, B, C listed in Table ~\ref{tab:table1}.
In Eq.(11) the radial scaling variable $X_F^R$ is given by
\begin{equation}
X_F^R=\frac{\sqrt{{P_l^*}^{2}+P_{\bot}^2}}{P_{max}^*},
\label{eq:11}
\end{equation}
where $P_l^*$ and $P_{\bot}$ are the longitudinal and transverse
momenta of pion in the pA center-of-mass system, respectively;
$P_{max}^*$ is the maximum value of $P^*$ allowed by the kinematics.
The expression (11)
reproduces all available data on small angle pion production in
proton-nucleus collisions in the projectile energy range from 5 to
24 GeV within the accuracy of about 25 \%. The $\pi^0$ spectrum
also needed for the calculations can be approximately expressed as
an arithmetic mean of the $\pi^+$ and $\pi^-$ spectra.

\begin{table}[htb!]
\caption{\label{tab:table1}Coefficients A, B, C of Eq.(11)}
\begin{ruledtabular}
\begin{tabular}{c|ccc|ccc}
 \multicolumn{1}{c|}{  }&\multicolumn{3}{c|}{Cu}&\multicolumn{3}{c}{Ta}\\
 \hline
   &  A   &  B   &  C    &  A   &   B   &     C\\
$\pi^+$ &  700 &  3.7 &  4.0  & 1000 &  3.8 &4.2 \\
$\pi^-$ &  380 &  4.2 &  4.7  &  650 &  3.9  & 4.5\\
\end{tabular}
\end{ruledtabular}
\end{table}

\section{\label{sec:level1}Data analysis }

For the present analysis the antiproton production in the
elementary $pN$ and $\pi N$ collisions  as well as the final
antiproton absorption in its way out of the nucleus are of
relevance. The parametrization of the total cross section for both
production channels is based on the results \cite{Ref19,Ref20}
obtained within the One-Boson-Exchange model:
\begin{equation}
  \sigma_{h N\to \bar pX}^{tot} = a(S/S_0-1)^b(S/S_0)^c. \label{eq:tot}
\end{equation}
Here $\sqrt S$ stands for the invariant collision energy, while
$\sqrt S_0$ denotes the threshold for the individual channel given
by the sum of the hadron masses in the final state of the
reaction. The values of the parameters $a$,~$b$,~$c$, used in the
following calculations, are listed in Table ~\ref{tab:table2}. The
analysis \cite{Ref19,Ref20} indicates a phase space dominance for
the antiproton production cross sections both from pion and
nucleon induced reactions.

\begin{table}[htb!]
\caption{\label{tab:table2}The parameters in the approximation
(13)}
\begin{ruledtabular}
\begin{tabular}{c|cccc}
 Reaction           &  $S_0$      &  $a$,~mb  & $b$       &  $c$   \\
 \hline
 $\pi N\to\bar pX$ & $9$  ~$m^2$ &     1   &    2.31   & -2.3\\
 $pN   \to\bar pX$ & $16$ ~$m^2$ &    0.12 &    3.5    & -2.7\\
 \hline
\end{tabular}
\end{ruledtabular}
\end{table}

The calculations by Eqs.~(\ref{eq:a}) - (\ref{eq:tot}), employing a
phase space behavior of the total $\bar p$  production cross
sections (Table ~\ref{tab:table2}) and a free inelastic $\bar pN$
cross section \cite{Ref25} as an effective in-medium
antiproton-nucleon cross section $\sigma_{\bar pN}^{eff}$,
essentially fail in describing the experimental data on Cu and Ta target nuclei.
The mean value of the ratio of the calculated and measured antiproton
production cross sections is equal to about 0.4.

Thus, in the next calculation we assume that for both antiproton
production channels the total elementary $\bar p$  production
cross sections follow the available phase space and treat
$\sigma_{\bar pN}^{eff}$ as a free parameter.
It is found that the experimental data can be reasonably reproduced
by the calculation implying $\sigma_{\bar pN}^{eff}=35$ mb for all
considered antiproton momenta (Fig.1). The uncertainty in the effective
in-medium antiproton-nucleon cross section caused by experimental errors 
is estimated to be $\pm 10$ mb.

\begin{figure}[t]
\includegraphics[height=0.76\linewidth]{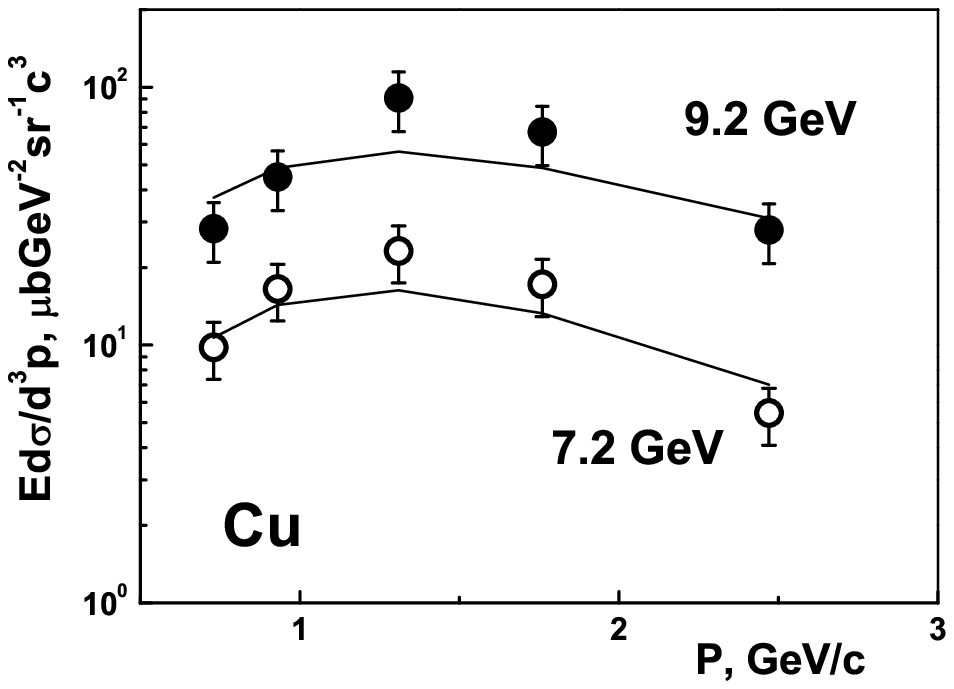}
\includegraphics[height=0.76\linewidth]{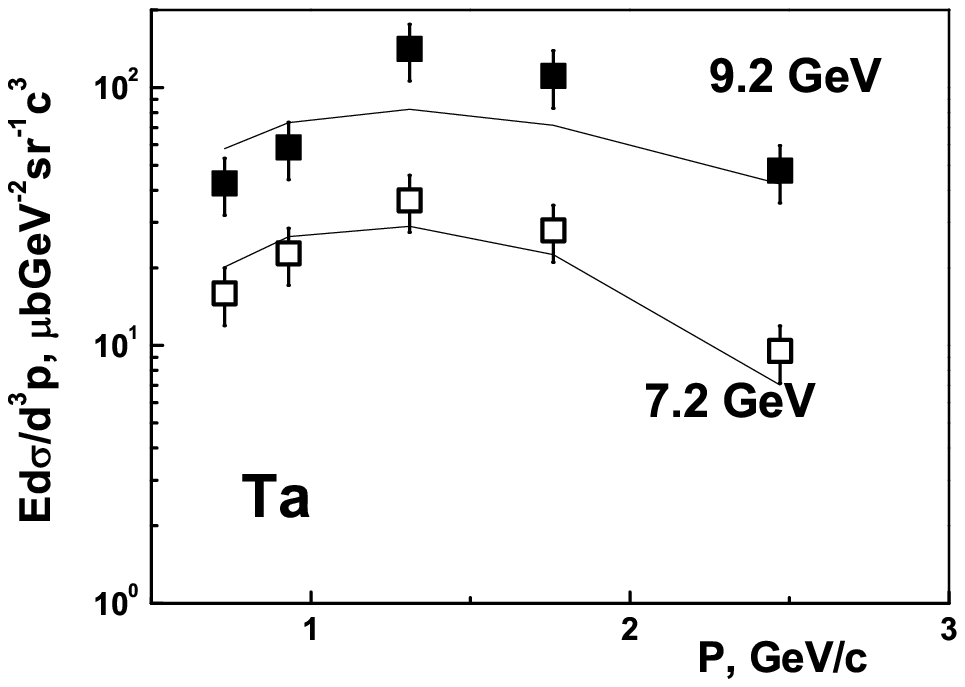}
\caption{\label{fig:fig1} Invariant cross sections for antiproton
production in $pCu$ and $pTa$ interactions at initial proton energies of
7.2 and 9.2 GeV as a function of $\bar p$ momentum. The experimental
data are from \cite{Ref18}. The curves are our calculations with
$\sigma_{\bar pN}^{eff} = 35 mb$}
\end{figure}

The calculations by Egs ~(\ref{eq:a}) - (\ref{eq:tot}) provide
the possibility to study the relative contributions of the direct
and cascade antiproton production mechanisms. It is found that the
strengths of these competing production channels slightly depend
on the effective in-medium $\bar pN$ cross section. The
contribution of the direct channel to the antiproton production
cross section at projectile proton energy of 7.2 GeV, defined as
a ratio $R= (\frac{E}{P^2}\frac{d^2\sigma}{dPd\Omega})^{dir}/[
(\frac{E}{P^2}\frac{d^2\sigma}{dPd\Omega})^{dir}+(\frac{E}{P^2}\frac{d^2\sigma}{dPd\Omega})^{casc}]$,
is shown in Fig.2 for Cu and Ta targets. One can see that the role
of the direct mechanism increases with antiproton momentum while
the role of the cascade mechanism decreases
as more energetic pions are required to produce fast antiproton.
Since the cascade mechanism probes more dense layers of the nucleus
the knowledge of these contributions is required for the estimate of
the nuclear density involved in the $\bar p$ production.

\begin{figure}[t]
\includegraphics[height=0.76\linewidth]{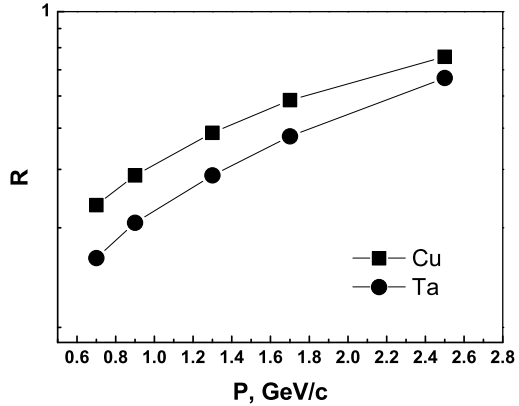}
\caption{\label{fig:fig2} Contribution of the direct channel to
the antiproton production cross section at incident proton energy
of 7.2 GeV as a function of $\bar p$ momentum.}
\end{figure}

Furthermore, it is well known that in the optical model the
particle absorption in nucleus is connected with the imaginary
part of the respective nuclear optical potential, which provides
the possibility to evaluate its magnitude.

The propagation through the nuclear matter of the $\phi$ mesons
produced in $\gamma$A \cite{Ref26} and $p$A \cite{Ref27,Ref28}
reactions has been studied within the Glauber type scattering
model. The change in the $\phi$ absorption was attributed to the
modification of its width in nuclear medium. Following the receipt
of Ref. \cite{Ref27} we express the antiproton survival
probability in the form:
\begin{equation}
 W=\exp(\int \limits_{z}^{\infty} dx Im \Pi(P_{\bar p},\rho(r'))/P_{\bar
 p}).
    \label{eq:13}
\end{equation}
Here $\Pi(P_{\bar p},\rho(r'))$ be the antiproton self-energy in a
nuclear medium as a function of its momentum $P_{\bar p}$ and the
nuclear density $\rho(r')$ with the antiproton production point
inside the nucleus $r^{'}=\sqrt{b^2+x^2}$.

Since the nuclear optical potential $U= \Pi(P_{\bar
p},\rho)/2E_{\bar p}$, Eq. (14) can be rewritten as:
\begin{equation}
    W=\exp(2\int \limits_{z}^{\infty} dx
    Im U(P_{\bar p},\rho(r'))/\beta).
    \label{eq:14}
\end{equation}
where $\beta = P_{\bar p}/E_{\bar p}$ stands for $\bar p$ velocity
in the target nucleus frame of reference.

On the other hand, this probability in Eq.(1) and Eq.(7) is read as:
\begin{equation}
    W=\exp(-\int \limits_{z}^{\infty} dx \sigma_{\bar pN}^{eff}A\rho(r')).
    \label{eq:15}
\end{equation}
From the comparison of Eq.(15) and Eq.(16) one gets the expression for
the imaginary part of the antiproton optical potential in the
nuclear matter:
\begin{equation}
   Im U = -\beta \sigma_{\bar pN}^{eff}A\rho(r')/2.
   \label{eq:16}
\end{equation}

Now we can evaluate the  $Im U$ at an estimated average nuclear
density involved $A\rho=0.36\rho_0$ assuming the mean value of the
$\sigma_{\bar pN}^{eff} =35$ mb. The imaginary part of the
antiproton optical potential changes from -14 MeV at antiproton
momentum of 0.73 GeV/c till -20 MeV at 2.5 GeV/c due to its
dependence on the antiproton velocity $\beta$. Thus, the magnitude
of  $Im U$ can be estimated as $-(38-56)$ MeV at $A\rho = \rho_0$.
These values are sizably larger than the imaginary part of the
antiproton nuclear optical potential derived from the analysis of
data on antiprotonic atoms \cite{Ref29} which is close to -160
MeV. However, in \cite{Ref29} the optical potential is determined
at radii where density is $(0.02-0.03) \rho_{0}$ and the
extrapolation of this potential to the central nuclear density is
highly model dependent. Note that our estimate of the imaginary
part of $\bar p$-nuclear potential is comparable with the
magnitudes of $ImU$ calculated for the $\phi$ mesons
\cite{Ref26,Ref27} as well as for the antikaons \cite{Ref30}.

\section{\label{sec:level1}Conclusion }

The analysis of the data on antiproton production on Cu and Ta nuclei within
the folding model, including proton- and pion-induced production
channels, evidences for the effective in-medium $\bar p$N cross section
in the momentum range 0.7-2.5 GeV/c is within $(25-45)$ mb. This
value differs from free total and inelastic $\bar p$N cross
sections, which demonstrate strong momentum dependence and
significantly larger magnitudes. The imaginary part of the
antiproton nuclear optical potential is estimated to be $-(38-56)$
MeV at normal nuclear matter density.

The authors acknowledge valuable discussions with V.A.Sheinkman
and for providing us with experimental data prior to publication.



\end{document}